\documentclass[aps,prc,reprint]{revtex4-1}
\usepackage{graphicx}
\usepackage{amsmath}

\begin{document}


\title{Statistical properties of thermal neutron capture cross section calculated with randomly generated resonance parameters} 



\author{N. Furutachi}
\author{F. Minato}
\author{O. Iwamoto}
\affiliation{Nuclar Data Center, Japan Atomic Energy Agency, Tokai-mura, Naka-gun, Ibaraki 319-1195, Japan}


\date{\today}

\begin{abstract} 
We investigated the probability distribution of the thermal neutron capture cross section
($\sigma_{\rm th}$) deduced stochastically with the resonance parameters randomly sampled 
from Wigner and Porter-Thomas distributions.
We found that the typical probability distribution has an asymmetric shape. 
While there is a long tail on the large $\sigma_{\rm th}$ side 
due to a resonance happening to be close to the thermal energy, 
the multi-resonance contribution considerably reduces the probability on the small $\sigma_{\rm th}$ side.  
We also found that the probability distributions have a similar shape if 
nuclei have an average resonance spacing sufficiently larger than an average radiation width.
We compared the typical probability distribution with the distribution of the 
experimental values of 193 nuclei, and found a good agreement between them. 
\end{abstract}


\maketitle 

\section{Introduction}
In a low-energy neutron reaction model, the compound nuclear process is usually described by resonance theories or 
the statistical model depending on the incident neutron energy.
Practically, resonance theories are used to fit experimental
cross sections and extract the resonance parameters.
In general, if there is no experimental data, 
the calculated cross sections using the resonance theories hardly make sense, 
because the theoretical prediction of  
the resonance parameter with high accuracy is extremely difficult due to the complexity of 
excited states of the compound nucleus around the neutron separation energy.
In a higher neutron incident energy region, 
where neighbor resonances overlap each other within their width, 
the resonance structure of the cross sections disappears.
In such region, cross sections can be characterized by the  
statistical average of the resonance parameters.

An approach to deduce cross sections in the resonance region 
from the statistical properties of the resonance parameters 
was made by Rochman $et$ $al$. \cite{1}. 
They generated resonance parameters by random sampling from the $\chi^2$ distribution for the 
neutron width, and Wigner distribution \cite{9} for the resonance energy spacing.
This approach is similar to the numerical method to account for the width fluctuation in 
the averaged cross section, which was performed by Moldauer \cite{4}.

The thermal neutron capture cross section $\sigma_{\rm th}$
is important for various nuclear applications and has been experimentally measured for most of stable nuclei.
Although $\sigma_{\rm th}$ data of radioactive nuclei are even required for particular applications, 
such as the evaluation of the transmutation system for the radioactive nuclear wastes, 
experimental data is limited compared with the stable nuclei. 
If there is no experimental data, $\sigma_{\rm th}$ is often calculated using the systematics \cite{2,3}.
However, $\sigma_{\rm th}$ is considered to be governed by the resonance that is closest to the 
thermal neutron energy i.e. the neutron separation energy of the 
compound nucleus, and drastically changed even with a slight variation of the resonance energy or width.
Therefore, the systematics would give only a rough estimation of $\sigma_{\rm th}$. 
Actually, they never predict extremely large $\sigma_{\rm th}$ observed for some nuclei, 
such as $^{135}$Xe (2.6 $\times$ $10^6$ b) and $^{157}$Gd (2.5 $\times$ $10^5$ b). 
If one applies the stochastic approach presented by Rochman {\it et al}. \cite{1}, 
such extremely large $\sigma_{\rm th}$ can happen to be obtained
if a resonance is generated at a energy close to the thermal energy.
Of course that does not necessarily happen, and 
the generated resonance sometimes appear far away from the thermal energy. 
Accordingly, such a calculated cross section must have a large uncertainty that originates from the statistical 
fluctuation of the resonance parameters.

Although such kind of stochastic approach will be an optional way to calculate the neutron reaction data 
in the resolved resonance region for nuclei that have no or scarce experimental data, 
it is important to clarify unavoidable uncertainty attributed to the statistical fluctuation at the same time.  
Therefore, in this study, we aim to figure out the uncertainty in $\sigma_{\rm th}$ predicted  
from the statistical property of the resonance parameters.
For this purpose, we deduced the probability distribution of $\sigma_{\rm th}$
from stochastically prepared sets of resonance parameters 
with their known statistical properties, and investigated its property. 
The probability distributions of the cross sections in the resolved resonance region 
deduced by such a stochastic method have been already utilized 
in the recent evaluated nuclear data library \cite{11}. 
To figure out the feature of the numerically obtained distribution, 
we also derive the analytical expression of the probability distribution by applying 
a single-resonance approximation, 
in which only one resonance closest to the neutron threshold is included.  

Practically, we calculate $\sigma_{\rm th}$ using Breit-Winger formula with the resonance parameters 
randomly sampled from Porter-Thomas \cite{10} and Wigner \cite{9} distributions.
The probability distribution of $\sigma_{\rm th}$ is numerically derived by 
calculating $\sigma_{\rm th}$ repeatedly using resonance parameter sets generated from different random seeds.
Porter-Thomas and Wigner distributions are characterized 
by the averages of the resonance widths and spacing, respectively,  
which are required as input of the present calculation.
We intend to apply the present 
method to the nuclei that have no experimental data in future, and in that case  
those parameters must be calculated theoretically. 
However, in this paper, we focus on nuclei whose $\sigma_{\rm th}$ and averages of the resonance spacing and widths
are known to confirm the validity of our method.

This paper is organized as follows. In Sec. \ref{theory}, numerical procedure to calculate 
the probability distribution of $\sigma_{\rm th}$ is explained, 
and the analytical expression of the probability distribution derived by applying the single-resonance approximation is presented. Properties of the calculated probability distribution of $\sigma_{\rm th}$ are discussed in Sec. \ref{res},  and Sec. \ref{sum} summarizes this work.

\section{Theoretical framework} \label{theory}

We use Breit-Wigner formula to calculate the s-wave radiative capture cross section at thermal energy $E_{\rm th}$,  
\begin{eqnarray} \label{eqbm}
\sigma_{\rm th} = \frac{\pi}{k^2}\sum_{j=I-\frac{1}{2}}^{j=I+\frac{1}{2}} \sum_{i \in j}^{N} \frac{g_j \Gamma_{ni} \Gamma_{\gamma i}}{\left( E_0-E_i \right)^2+\left( \frac{1}{2} \Gamma_{i} \right)^2} .
\end{eqnarray}
In this equation, $\Gamma_{ni}, \Gamma_{\gamma i}, \Gamma_{i},$ $E_i$, and $N$ are the neutron width, the radiative width, the total width, the energy of the i-th resonance with total spin $j$, and the number of resonances, respectively.
Fission, alpha and other particle emission channels are not considered in the present formulation. 
The wave number $k=\sqrt{2\mu E_0}/\hbar$ and $E_0=m_t/\left( m_n+m_t \right) E_{{\rm th}}$, where $m_n, m_t,$ and $\mu$ are the neutron mass, the target mass and the reduced mass, respectively. The spin statistical factor $g_j$ is given by
\begin{eqnarray}
g_j=\frac{2j+1}{2(2I+1)},
\end{eqnarray}
where $I$ is the total spin of the target nucleus.
In our framework, the resonance energy is a random variable given by Wigner distribution \cite{9}, 
\begin{eqnarray}\label{eqw}
P_{\rm e}(x^j)=\frac{\pi}{2}x^j e^{-\frac{\pi}{4}(x^j)^2}, x^j = \frac{E_{i+1}^j-E_i^j}{D_0^j}, 
\end{eqnarray}
with the average s-wave resonance spacing $D_0^j$. 
Since the experimental value of $D_0$ does not depend on $j$, 
we evaluate $D_0^j$ of $I\neq0$ nuclei from 
the experimental $D_0$ by adopting Gaussian spin distribution with the spin dispersion parameter ${\rm s}^2$ of 
Mengoni-Nakajima \cite{7},  
\begin{eqnarray}\label{eq_spindist}
W(E_x,j) &=& \frac{2j+1}{2{\rm s}^2}\exp\left[ -\frac{(j+\frac{1}{2})^2}{2{\rm s}^2}\right].
\end{eqnarray}
Using Eq. \eqref{eq_spindist}, $D_0^j$ are calculated as
\begin{eqnarray}
D_0^{I+\frac{1}{2}} &=& \frac{1+R_I}{R_I} D_0, \hspace{2mm} D_0^{I-\frac{1}{2}} = (1+R_I) D_0, \\
R_I &=& \frac{W(S_n,I+\frac{1}{2})}{W(S_n,I-\frac{1}{2})}. \nonumber
\end{eqnarray}

The neutron width is related with the reduced neutron width as
\begin{eqnarray}
\Gamma_{ni}=\sqrt{E_0}\Gamma_{ni}^{l=0}, 
\end{eqnarray}
and the reduced neutron width is a random variable with Porter-Thomas distribution \cite{9},
\begin{eqnarray}\label{eqp}
P_{\rm w} \left( \Gamma_{ni}^{l=0}\right) =\frac{1}{\sqrt{2\pi\Gamma_{ni}^{l=0} \langle \Gamma_{n0} \rangle}} \exp \left( -\frac{\Gamma_{ni}^{l=0}}{2\langle \Gamma_{n0} \rangle} \right). 
\end{eqnarray}
We calculate the average s-wave neutron reduced width $\langle \Gamma_{n0} \rangle$ using the experimental strength function
$S_0$ defined as
\begin{eqnarray} \label{eq10}
S_0=\frac{\langle g\Gamma_{n0}\rangle}{D_0}=\frac{1}{\Delta E}\sum_j \sum_{i \in j} g_j \Gamma_{ni}^{l=0}.
\end{eqnarray} 
For $I=0$ nucleus, $\langle \Gamma_{n0}\rangle = S_0 D_0$. For $I \neq 0$ nucleus, we suppose that  
$\Gamma_{ni}^{l=0}$ is not dependent on $j$, and write Eq. \eqref{eq10} as   
\begin{eqnarray} \label{eq11}
S_0 = \frac{1}{D_0}  \left( g_{I-\frac{1}{2}} \langle \Gamma_{n0}\rangle \frac{D_0}{D_0^{I-\frac{1}{2}}} 
+ g_{I+\frac{1}{2}} \langle \Gamma_{n0}\rangle \frac{D_0}{D_0^{I+\frac{1}{2}}} \right).
\end{eqnarray}
The average neutron width  $\langle \Gamma_{n0}\rangle$ is calculated using Eq. \eqref{eq11} with the experimental $S_0$. 
For $\Gamma_{\gamma i}$, we assume that 
\begin{eqnarray}
\Gamma_{\gamma i}\sim \langle \Gamma_{\gamma 0} \rangle.
\end{eqnarray}

In the calculation of actual nuclei, we use the experimental data of $D_0$, 
$\langle \Gamma_{\gamma 0} \rangle$ and $S_0$, which
are taken from the compilation of Mughabghab \cite{5}.
We calculate the probability distribution of 193 nuclei for which all of 
the resonance parameters and $\sigma_{\rm th}$ are given by Mughabghab. 
Since the fission channel is not taken into account in the present formulation,
the nuclei heavier than $^{209}$Bi are excluded.
We also arbitrarily exclude the nuclei lighter than $^{27}$Al
because it is expected that the statistical property of the resonance parameter is 
invalid for nuclei with low level densities.

As explained above, the resonance parameters are random variables in the present framework,
and therefore, $\sigma_{\rm th}$ has a probability distribution. 
We calculate it in two ways; one is a numerical way that is the repetitive calculation of $\sigma_{\rm th}$ 
using randomly sampled resonance parameters (Sec. \ref{numerical}),
and another is an analytical way using the single-resonance approximation (Sec. \ref{analytical}). 
  
\subsection{Numerical solution} \label{numerical}

First of all, we set an initial resonance at $-\frac{N}{2}\times D_0$ so that it starts from a 
sufficiently small energy. Then the next resonance energy is randomly sampled from Winger distribution of Eq. \eqref{eqw}. 
For each resonance, the width is randomly sampled
from Porter-Thomas distribution of Eq. \eqref{eqp}. For $I\neq 0$ nucleus, $j=I-\frac{1}{2}$ and $j=I+\frac{1}{2}$ 
resonances are generated independently.
The generation of the resonance is repeated until the number of resonance reaches to $N$, 
and then we calculate ${\sigma_{\rm th}}$ using the set of the resonance parameters and Breit-Wigner formula.
By repeating this procedure, we obtain many samples of $\sigma_{\rm th}$.
With sufficiently large number of samples, the probability distribution of $\sigma_{\rm th}$, 
which is denoted by $P_{\sigma}^{\rm M}$ (M stands for multi-resonance), is obtained.

To discuss the statistical property of $P_{\sigma}^{\rm M}$, 
we calculate the median $\sigma_{\rm md}^{M}$ and the dispersion $V$, which is defined as   
\begin{eqnarray}  \label{disp}
V = \int_{-\infty}^{\infty} ( x-\log \sigma_{\rm md}^{\rm M} )^2 P_{\sigma}^{\rm M}(x) dx, \hspace{3mm} x = \log \sigma_{\rm th}. 
\end{eqnarray}
It is noted that the definition is different from the usual one of the standard deviation.
The dispersion $V$ is also used to check the convergence of the numerical procedure.
If the number of trials is small, $V$ values obtained from different random seeds 
are significantly different from each other because of the statistical fluctuation. 
We fixed the number of trials at $10^7$ to obtain $P_{\sigma}^{\rm M}$ for most calculations in this paper.
Then the variation of $V$ caused by the statistical fluctuation is estimated to be less than 0.5$\%$.

It is noted that $P_{\sigma}^{\rm M}$ is dependent on the number of resonances $N$. 
We used $N=50$ in this paper, and confirmed that $V$ calculated using $N=100$ 
has less than 1$\%$ difference with that using $N=50$ in the calculation of $^{120}$Sn 
shown in Sec. \ref{res}. The dependence of $P_{\sigma}^{\rm M}$ on $N$ is also discussed there.

\subsection{Analytical solution with single-resonance approximation} \label{analytical}
To understand the property of $\sigma_{\rm th}$, 
we present the formulation with the single-resonance approximation. 
This approximation enables us to express the probability distribution of 
$\sigma_{\rm th}$ in an analytical form and provides us a deeper insight into its properties.
 
Supposing that $\sigma_{\rm th}$ is dominated by the first resonance, and
the thermal energy and the total width are small compared to the resonance energy, 
Eq. \eqref{eqbm} is approximated as
\begin{eqnarray} \label{eqa}
\sigma_{\rm th} &\simeq& \frac{\pi}{k^2} \frac{g_j \Gamma_{n1} \Gamma_{\gamma 1}}{E_1^2} , \\
&\propto& \frac{\Gamma_{n1}^{l=0}}{E_1^2}. \nonumber
\end{eqnarray}
Here $E_1$ is the energy of the resonance that is closest to $E_0$. 
Note that both positive and negative resonances can be $E_1$.
In this equation, $E_1$ and $\Gamma_{n1}^{l=0}$ are random variables.

We found that if the resonance energy spacing is given by Wigner distribution, the probability distribution 
of $E_1$ can be given by the normal distribution with the standard deviation of $d=D_0^2/2\pi$ for $I=0$ nucleus.
For $I\neq 0$ nuclei, 
we approximate $E_1$ with the normal distribution 
same with $I=0$ case using $j$-independent $D_0$.
   
Since the probability distribution of $E_1$ is the normal distribution, that of  
$E_1^2$ is the $\chi^2$ distribution of one degree of freedom.
Then $\sigma_{\rm th}$ is found to be proportional to the ratio of two $\chi^2$ distributions with 
Porther-Thomas distribution, which is generally known to be given by F-distribution.
As a consequence of the above explained conversions of the statistical variables (the detail is given in Appendix), 
the probability distribution of $\sigma_{\rm th}$ is given as 
\begin{eqnarray}\label{eqf}
 P_{\sigma}^{\rm S} (\sigma_{\rm th}) &=& \frac{1}{\pi \sigma_{\rm th}}\left( \sqrt{\frac{\sigma_{\rm th}}{\sigma_{{\rm md}}^{\rm S}}}+\sqrt{\frac{\sigma_{{\rm md}}^{\rm S}}{\sigma_{\rm th}}} \right)^{-1}, \\
 \sigma_{{\rm md}}^{{\rm S}} &=& \frac{2 \pi^2}{k^2} \sqrt{E_0} \frac{\langle \Gamma_{n0} \rangle \langle \Gamma_{\gamma 0} \rangle}{D_0^2},
\end{eqnarray}
where $\sigma_{{\rm md}}^{{\rm S}}$ corresponds to the median of the probability distribution (S stands for single resonance).
Since the expectation value of $\sigma_{\rm th}$ is dominated by the tail of the probability  
distribution in extremely large $\sigma_{\rm th}$ region, 
we use the median, not the average, in what follows.

We express $P_{\sigma}^{\rm S}$ as well as $P_{\sigma}^{\rm M}$ 
in terms of log $\sigma_{\rm th}$ instead of $\sigma_{\rm th}$.
It was found that $P_{\sigma}^{\rm S}(\log \sigma_{\rm th})$ is expressed by a hyperbolic secant function 
centered at $\log \sigma_{{\rm md}}^{{\rm S}}$,
\begin{eqnarray}\label{eqPf}
P_{\sigma}^{\rm S}(\log \sigma_{\rm th}) = \frac{1}{2\pi} {\rm sech}\left( \frac{\log \sigma_{\rm th} - \log \sigma_{{\rm md}}^{{\rm S}}}{2} \right).
\end{eqnarray} 
It is clear from this equation that the averages of the resonance spacing and widths, which are characteristic of nuclei, changes only the center of the probability distribution. In other words, $P_{\sigma}^{\rm S}(\log\frac{\sigma_{\rm th}}{\sigma_{\rm md}^{\rm S}})$ is identical for all nuclei.

\section{results} \label{res}

\subsection{Probability distributions for specific nuclei}
 
\begin{figure}[h]
	\includegraphics[scale=0.35,angle=270]{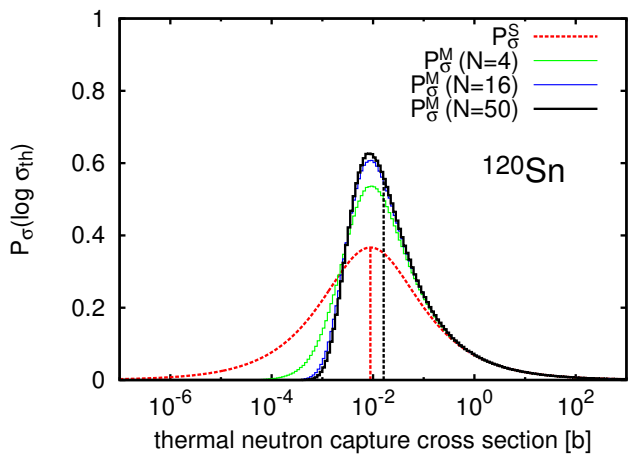}
	\caption{ Probability distributions $P_{\sigma}^{\rm M}(\log \sigma_{\rm th})$ and $P_{\sigma}^{\rm S}(\log \sigma_{\rm th})$. For $P_{\sigma}^{\rm M}(\log \sigma_{\rm th})$, calculations using the number of resonance $N$=4,16,50 are compared. } \label{fig2}
\end{figure}

In this section, we show the probability distributions 
$P_{\sigma}^{\rm M}$ and $P_{\sigma}^{\rm S}$ for specific nuclei. 
It is noted that $P_{\sigma}^{\rm S}$ is essentially different from $P_{\sigma}^{\rm M}$ 
due the absence of the multi-resonance contribution. 
In Fig. \ref{fig2}, $P_{\sigma}^{\rm M}$ and $P_{\sigma}^{\rm S}$ as function of $\log \sigma_{\rm th}$ 
calculated using the experimental values of the averages of the resonance spacing and widths are shown.
First of all, it is clearly seen that 
both $P_{\sigma}^{\rm M}(\log \sigma_{\rm th})$ and $P_{\sigma}^{\rm S}(\log \sigma_{\rm th})$ 
are distributed over extremely wide range of several order of magnitude.
This broad distribution is naturally regarded from the property of resonance nature. 

As for $P_{\sigma}^{\rm S}(\log \sigma_{\rm th})$, it is just a hyperbolic secant function, 
which has a symmetric shape and long tails both in small and large $\sigma_{\rm th}$ regions.
The peak of the distribution corresponds to $\sigma_{{\rm md}}^{\rm S}$, which is 0.01 b for $^{120}$Sn.  
While we chose $^{120}$Sn as an example here, 
the shape of $P_{\sigma}^{\rm S}(\log \sigma_{\rm th})$ is the same for all nuclei, as mentioned in 
the last paragraph of the Sec. \ref{analytical}.

Although it is expected that $\sigma_{\rm th}$ is dominated by the first resonance,
$P_{\sigma}^{\rm M}(\log \sigma_{\rm th})$ shows a considerably 
different distribution from $P_{\sigma}^{\rm S}(\log \sigma_{\rm th})$.
To show the dependence on the number of resonance $N$, 
$P_{\sigma}^{\rm M}$ calculated using $N=4,16,50$ are compared in Fig. \ref{fig2}.
While the tail of $P_{\sigma}^{\rm S}(\log \sigma_{\rm th})$ extends to  
the small cross section side, $P_{\sigma}^{\rm M}(\log \sigma_{\rm th})$ has a steep slope on the left shoulder.
Comparing the calculations with different $N$,
it can be seen $P_{\sigma}^{\rm M}$ with the larger $N$ has a less probability in the smaller $\sigma_{\rm th}$ side.
On the other hand, $P_{\sigma}^{\rm M}(\log \sigma_{\rm th})$ and $P_{\sigma}^{\rm S}(\log \sigma_{\rm th})$ 
are similar in the larger $\sigma_{\rm th}$ region. 
They have peaks at approximately the same $\sigma_{\rm th}$, and overlaps in the region of $\sigma_{\rm th} > 1$ b.
In $P_{\sigma}^{\rm M}(\log \sigma_{\rm th})$, the second and subsequent resonances 
have considerable contributions to $\sigma_{\rm th}$ if its value is small, 
while the first resonance has a dominant contribution to the emergence of a large $\sigma_{\rm th}$.
Consequently, the multi-resonance contribution considerably reduces 
the statistical fluctuation in small values of $\sigma_{\rm th}$.

%
%

\begin{figure}[h]
	\includegraphics[scale=0.41,angle=270]{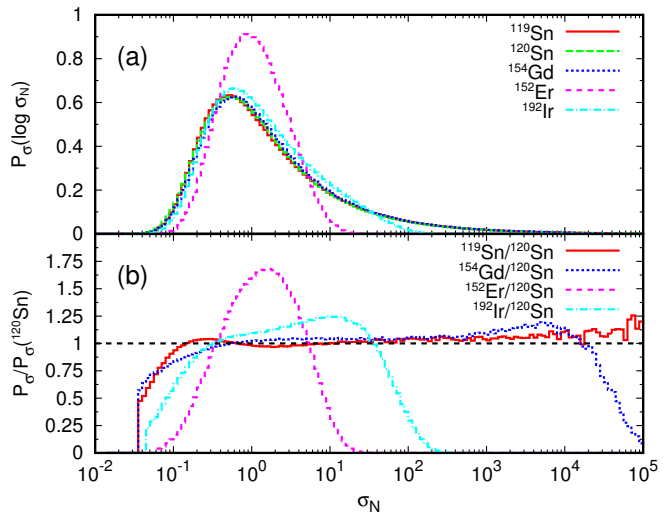} \\
	\caption{ (a) Probability distribution $P_{\sigma}^{\rm M}(\log \sigma_{\rm N})$ 
of $^{119}$Sn, $^{120}$Sn, $^{152}$Eu, $^{154}$Gd, and (b) the ratio of those to $P_{\sigma}^{\rm M}(\log \sigma_{\rm N})$ of $^{120}$Sn.} \label{fig3}
\end{figure}

We also calculated $P_{\sigma}^{\rm M}$ of various nuclei, and
found that the difference of the averages of the resonance spacing and widths barely affect 
the shape of $P_{\sigma}^{\rm M}(\log \sigma_{\rm th})$, 
if the condition $D_0 \gg \langle \Gamma_{\gamma 0} \rangle$ is satisfied. 
To show this, we compare shapes of $P_{\sigma}^{\rm M}(\log \sigma_{\rm th})$ 
for different nuclei by normalizing $\sigma_{\rm th}$ to medians of distributions.  
For simplicity, we use the following notation 
\begin{eqnarray}
\sigma_{\rm N}=\frac{\sigma_{\rm th}}{\sigma_{{\rm md}}}.
\end{eqnarray} 
In Fig. \ref{fig3} (a), $P_{\sigma}^{\rm M} (\log \sigma_{\rm N})$ of several nuclei are 
shown as a function of $\sigma_{\rm N}$. 
We also show the ratio to the $P_{\sigma}^{\rm M} (\log \sigma_{\rm N})$ 
of $^{120}$Sn in Fig. \ref{fig3} (b).     
Comparing the results of $^{120}$Sn and $^{154}$Gd, 
which have $D_0=1485$ eV and 13.8 eV, respectively, 
$P_{\sigma}^{\rm M}(\log \sigma_{\rm N})$ are very similar in spite of the large difference of $D_0$.
Only a small difference is found in $\sigma_N<1$ and $\sigma_N>10^4$ as seen in Fig. \ref{fig3} (b).
For most of the calculated 193 nuclei, the differences are even smaller than this case or the same degree.
Exception is found in  $^{152}$Eu and $^{192}$Ir, 
which have extremely small $D_0$ of 0.25 eV and 0.64 eV, respectively.
The distributions of $^{152}$Eu and $^{192}$Ir are  
significantly narrower range than the typical distribution represented by that of $^{120}$Sn.
In Sec. \ref{rpdep}, we show that such noticeable difference appears 
if nuclei have $D_0$ comparable to $\langle \Gamma_{\gamma 0} \rangle$.

Differences between $P_{\sigma}^{\rm M}(\log \sigma_{\rm N})$ for $I=0$ and $I\neq$ 0 nuclei are also extremely small. 
Comparing $P_{\sigma}^{\rm M}(\log \sigma_{\rm N})$ of $^{119}$Sn ($I=1/2$) and $^{120}$Sn ($I=0$), 
only a minute difference is found around $\sigma_N \sim 0.1$.

\subsection{Dependence on averages of resonance spacing and widths} \label{rpdep}

We investigated the dependence of $P_{\sigma}^{\rm M} (\log \sigma_{\rm N})$ 
on the averages of the resonance spacing and widths.
To quantify a change of $P_{\sigma}^{\rm M} (\log \sigma_{\rm N})$, 
the dispersion $V$ defined by Eq.\eqref{disp} is discussed in this section.
Further discussion using the quantities with higher-moments is given in Sec. \ref{rpdep_detail}. 
We also discuss $\sigma_{{\rm md}}^{\rm M} $ normalized to $\sigma_{{\rm md}}^{\rm S}$ ($\sigma_{{\rm md}}^{\rm M}/\sigma_{{\rm md}}^{\rm S} $).
By this quantity, we can measure the change of $P_{\sigma}^{\rm M}$ relative to $P_{\sigma}^{\rm S}$, 
which is independent of the averages of the resonance spacing and widths.
If  $\sigma_{{\rm md}}^{\rm M}/\sigma_{{\rm md}}^{\rm S} $ is smaller (lager) than 1, 
it means that $P_{\sigma}^{\rm M}$ concentrated in the small (large) cross section side relative to $P_{\sigma}^{\rm S}$. 

\begin{figure*} [t]
	\includegraphics[scale=0.8,angle=270]{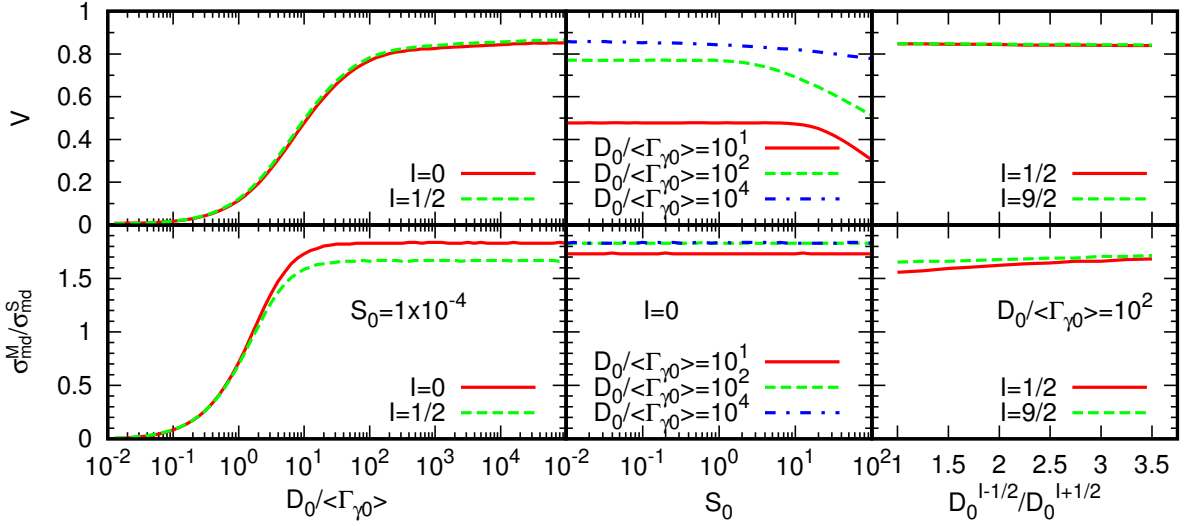}
	\caption{ Dependence of $P_{\sigma}^{\rm M} (\log \sigma_{\rm N})$ on the averages of the resonance spacing, widths and the target spin $I$. Calculated $\sigma_{{\rm md}}^{\rm M}/\sigma_{{\rm md}}^{\rm S}$ and $V$ as functions of $D_0/\langle \Gamma_{\gamma 0} \rangle$, $S_0$ and $D_0^{I-1/2}/D_0^{I+1/2}$ are shown in the top and bottom panels, respectively. Results shown in the left, middle and right panels are obtained using  
$S_0=1$, $I$=0 and $D_0/\langle \Gamma_{\gamma 0} \rangle$=100, respectively.
In all calculations, $\langle \Gamma_{\gamma 0}\rangle$ is fixed at 0.1 eV. } \label{dgdist}  
\end{figure*}

We calculated $P_{\sigma}^{\rm M}$ using various $D_0$ and $S_0$ for the target spin $I$=0 and $I=1/2$ cases.  
To see how the $j$-dependence of $D_0$ affects $P_{\sigma}^{\rm M}$, 
the ratio $D_0^{I-1/2}/D_0^{I+1/2}$ is also varied.
In Fig. \ref{dgdist}, the calculated $\sigma_{{\rm md}}^{\rm M}/\sigma_{{\rm md}}^{\rm S}$ and $V$ 
as a function of the averages of the resonance spacing and widths are shown.  
The calculation is carried out by fixing $\langle \Gamma_{\gamma 0} \rangle =$0.1 eV. 
We have confirmed that the result shown in Fig. \ref{dgdist} does not change 
even if $\langle \Gamma_{\gamma 0} \rangle=$1 eV.
  
In the left panels of Fig. \ref{dgdist}, the results with $S_0$ fixed at 1$\times 10^{-4}$ 
are shown as a function of $D_0/\langle \Gamma_{\gamma 0} \rangle$. 
In $D_0/\langle \Gamma_{\gamma 0} \rangle > 100$ region,
$\sigma_{{\rm md}}^{\rm M}/\sigma_{{\rm md}}^{\rm S}$ 
have almost constant values of 1.8 and 1.7 for $I=0$ and $I=1/2$ cases, respectively. 
Similarly, $V$ is almost converged at 0.8 in this region. 
In $D_0/\langle \Gamma_{\gamma 0} \rangle < 100$ region, a significant decrease of $V$ is seen.
Around $D_0/\langle \Gamma_{\gamma 0} \rangle \sim 100$, where $V$ starts to decrease, 
the tail of $P_{\sigma}^{\rm M}$ in the large $\sigma_{\rm th}$ side is slightly reduced as in the case of  $^{192}$Ir.
The median is rather insensitive to the variation of the tail distribution, 
therefore $\sigma_{{\rm md}}^{\rm M}/\sigma_{{\rm md}}^{\rm S}$ is still constant 
around $D_0/\langle \Gamma_{\gamma 0} \rangle \sim 100$.
If $D_0$ becomes comparable to $\langle \Gamma_{\gamma 0} \rangle$, 
$P_{\sigma}^{\rm M} $ distributes in 
a significantly narrower $\sigma_{\rm th}$ range, as in the case of $^{152}$Eu.

The middle panels of Fig. $\ref{dgdist}$ show the results as a function of $S_0$ with $I=0$. 
If $\langle \Gamma_{n0} \rangle$ is much lager than $\langle \Gamma_{\gamma 0} \rangle$, a total width 
is often dominated by a neutron width. In this case, $P_{\sigma}^{\rm M} (\log \sigma_{\rm N})$ changes 
from the typical distribution.
This can be seen from the middle top panel, in which the reduction of $V$ becomes noticeable in the large $S_0$ region. 
The degree of the change is not so large, 
therefore $\sigma_{{\rm md}}^{\rm M}/\sigma_{{\rm md}}^{\rm S}$ shows a constancy against the change of $S_0$.

Although $P_{\sigma}^{\rm M}$ is also dependent on $I$, it has an small influence 
on the shape of the distribution, but slightly change the median of the distribution. 
From the left bottom panel of Fig.\ref{dgdist}, the difference 
between $\sigma_{{\rm md}}^{\rm M}/\sigma_{{\rm md}}^{\rm S}$ calculated with 
$I=0$ and $I=1/2$ can be seen. 
The difference between the results with $I=1/2$ and $I=9/2$ is also visible in the right bottom panel.
These differences are mainly because $j$-independent $D_0$ 
and the spin statistical factor $g_j=$ 1 are assumed 
in the calculation of $\sigma_{\rm md}^{\rm S}$ if $I$ is not equal to 0.           
Therefore, the variation of $\sigma_{{\rm md}}^{\rm M}/\sigma_{{\rm md}}^{\rm S}$ is largest if the nucleus has $I=1/2$, because the difference between $g_{I-\frac{1}{2}}$ and $g_{I+\frac{1}{2}}$ is largest in this case. 
On the other hand, there is an extremely small change in $V$ calculated using different $I$, 
as shown in the left top and the right top panel.  

\begin{figure}[h]
	\includegraphics[scale=0.35,angle=270]{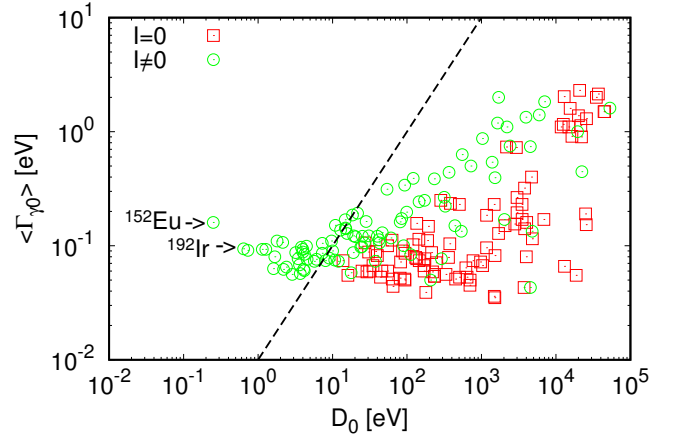}
	\caption{Distribution of $D_0$ and $\langle \Gamma_{\gamma 0} \rangle$ for 193 nuclei with $I=0$ and $I\neq0$. The dashed line indicates $D_0/ \langle \Gamma_{\gamma 0} \rangle = 100$ line. The symbols corresponding to $^{152}$Eu and $^{192}$In are indicated by the arrows.} \label{fig7}
\end{figure} 

As discussed above, one of the conditions that cause a significant change in $P_{\sigma}^{\rm M} (\log \sigma_{\rm N})$ is that 
$D_0$ becomes comparable to $\langle \Gamma_{\gamma 0} \rangle$.
To see whether actual nuclei satisfy this condition or not,
the experimentally known $D_0$ and $\langle \Gamma_{\gamma 0} \rangle$ of 193 stable nuclei are plotted in Fig. \ref{fig7}.
We can see that the most of nuclei are in the region of $D_0/\langle \Gamma_{\gamma 0} \rangle > 100$,
where $\sigma_{{\rm md}}^{\rm M}/\sigma_{{\rm md}}^{\rm S}$ and $V$ are almost constant.
Namely, the most of stable nuclei have very similar $P_{\sigma}^{\rm M}  $, which are characterized with 
$\sigma_{{\rm md}}^{\rm M}/\sigma_{{\rm md}}^{\rm S}\sim1.8$ (I=0) or 1.7 (I$\neq$0) and $V \sim 0.8$.
The constancy of $\sigma_{{\rm md}}^{\rm M}/\sigma_{{\rm md}}^{\rm S} $ will be convenient for a practical use, 
because $\sigma_{{\rm md}}^{\rm M}$ can be calculated easily from $\sigma_{{\rm md}}^{\rm S}$ 
without numerically calculating $P_{\sigma}^{\rm M}$ from many samples. 
In $D_0/\langle \Gamma_{\gamma 0} \rangle < 100$ side, there are approximately 30 nuclei.
Among these nuclei, $^{152}$Eu has the smallest $D_0/\langle \Gamma_{\gamma 0} \rangle$ of 1.56. 
While only $^{152}$Eu have extremely large $D_0$ comparable to $\langle \Gamma_{\gamma 0} \rangle$ among stable nuclei, 
it may not be a case for unstable nuclei. For example, since neutron separation energies of proton-rich nuclei are generally 
higher than those of stable nuclei, they may have extremely large $D_0$. 

\subsection{General properties of probability distributions} \label{rpdep_detail}

In this section, the general property of the probability distribution is discussed in terms of  
the skewness $\beta_1^{1/2}$ and the kurtosis $\beta_2$, 
which are defined by third and fourth central moments with the expectation value $\mu$, respectively,
\begin{eqnarray}
\mu &=&\int_{-\infty}^{\infty} x P_{\sigma}^{\rm M}(x) dx, \hspace{5mm} \nonumber \\ 
\mu_r &=& \int_{-\infty}^{\infty} ( x-\mu )^r P_{\sigma}^{\rm M}(x) dx, \hspace{5mm} x = \log \sigma_{\rm th}, \nonumber \\
\beta_1^{1/2} &=& \frac{\mu_3}{\mu_2^{3/2}}, \hspace{5mm} \beta_2=\frac{\mu_4}{\mu_2^2}-3.
\end{eqnarray}
The skewness $\beta_1^{1/2}$ has a sensitivity to the asymmetry of the distribution.  
The kurtosis is more sensitive to the tailedness of the distribution than the quantities with lower central moments.
Both quantities become 0 for the normal distribution in the present definition.

\begin{figure} 
\includegraphics[scale=0.35,angle=270]{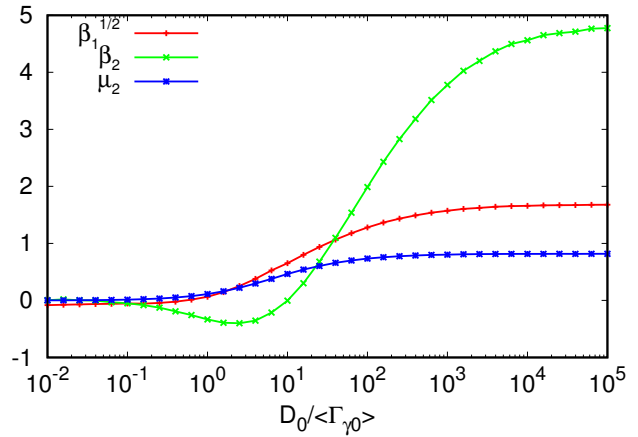}
\caption{Skewness $\beta_1^{1/2}$ and kurtosis $\beta_2$ of $P_{\sigma}^{\rm M}$ as a function 
of $D_0/\langle \Gamma_{\gamma 0} \rangle$.} \label{skewness}
\end{figure}

In Fig. \ref{skewness}, the calculated $\beta_1^{1/2}$, $\beta_2$ and $\mu^2$ are shown. 
In $D_0/\langle \Gamma_{\gamma 0} \rangle > 100$ region, while $\mu^2$ comes close to the constant, 
$\beta_1^{1/2}$ and $\beta_2$ are still increasing, 
which indicates that the tail of the distribution is growing in this region.
The positive sign of $\beta_1^{1/2}$ means that the distribution is leaning to the left side.
As already discussed in the previous section, the distribution drastically changes around 
$D_0/\langle \Gamma_{\gamma 0} \rangle \sim 10$, and $\beta_2$ becomes negative there.
The negative sign of $\beta_2$ means that the tail of the distribution is even suppressed 
from that of the normal distribution.

In $D_0/\langle \Gamma_{\gamma 0} \rangle < 1$ region, 
all quantities come close to 0, and the distribution converges to the normal distribution. 
In that condition, the following expressions may be valid.
Supposing $E_i^2 \ll 1/4\Gamma_i^2$ and $\Gamma_{\gamma i} \gg \Gamma_{ni}$ in Eq. \eqref{eqbm},
$\sigma_{\rm th}$ is approximated as
\begin{eqnarray}
\sigma_{\rm th} \simeq \frac{\pi}{k^2}\frac{\langle \Gamma_{\gamma 0} \rangle}{E_0^2+\frac{1}{4}\langle \Gamma_{\gamma 0} \rangle^2} \sum_{i}^{N_c} \Gamma_{ni},
\end{eqnarray}
where $N_c$ is the number of the last resonance which satisfies the above conditions.
If $N_c$ is sufficiently large, the probability distribution of $\sigma_{\rm th}$
is approximated with the normal distribution from the central limit theorem,
\begin{eqnarray}
P_{\sigma}^M (\sigma_{\rm th}) &\simeq& \frac{1}{\sqrt{2\pi \mu_2}}\exp \left( -\frac{(\sigma_{\rm th}-\mu)^2}{2\mu_2} \right).
\end{eqnarray}
Here $\mu$ and $\mu_2$ are given as
\begin{eqnarray}
\mu &=&  \frac{\pi}{k^2}\frac{\langle \Gamma_{\gamma 0} \rangle }{E_0^2+\frac{1}{4}\langle \Gamma_{\gamma 0} \rangle^2} \sqrt{E_0}  \langle \Gamma_{n0} \rangle N_c, \hspace{2mm}  \mu_2=\mu^2 \frac{2}{N_c}.
\end{eqnarray}
From these equations, it is clear that the standard deviation of the normal distribution 
$\mu_2$ decreases as $N_c$ increases. 
The uncertainty arising from the statistical fluctuation of the resonance parameters is minimized in this case.

\subsection{Comparison with the distribution of experimental data}

Finally, we compare $P_{\sigma}^{\rm M}$ with the experiments,
to confirm the validity of the present method.  
In principle, we cannot discuss the validity of $P_{\sigma}^{\rm M}$ for each nucleus, 
because there is only one experimental value for each nucleus to be compared with.
Therefore, we utilize the finding that $P_{\sigma}^{\rm M}(\log \sigma_{\rm N})$ are similar for most of nuclei. 
We normalize the experimental $\sigma_{\rm th}$ of 193 stable nuclei to the calculated $\sigma_{{\rm md}}^{\rm M}$, 
in order to compare them with the typical $P_{\sigma}^{\rm M}(\log \sigma_{\rm N})$ calculated 
with the condition of $D_0/\langle \Gamma_{\gamma 0} \rangle = 10^3$.
Namely, we check the validity of one typical $P_{\sigma}^{\rm M}(\log \sigma_{\rm N})$ 
by comparing it with the distribution of 193 samples.

\begin{figure}[h]
	\includegraphics[scale=0.35,angle=270]{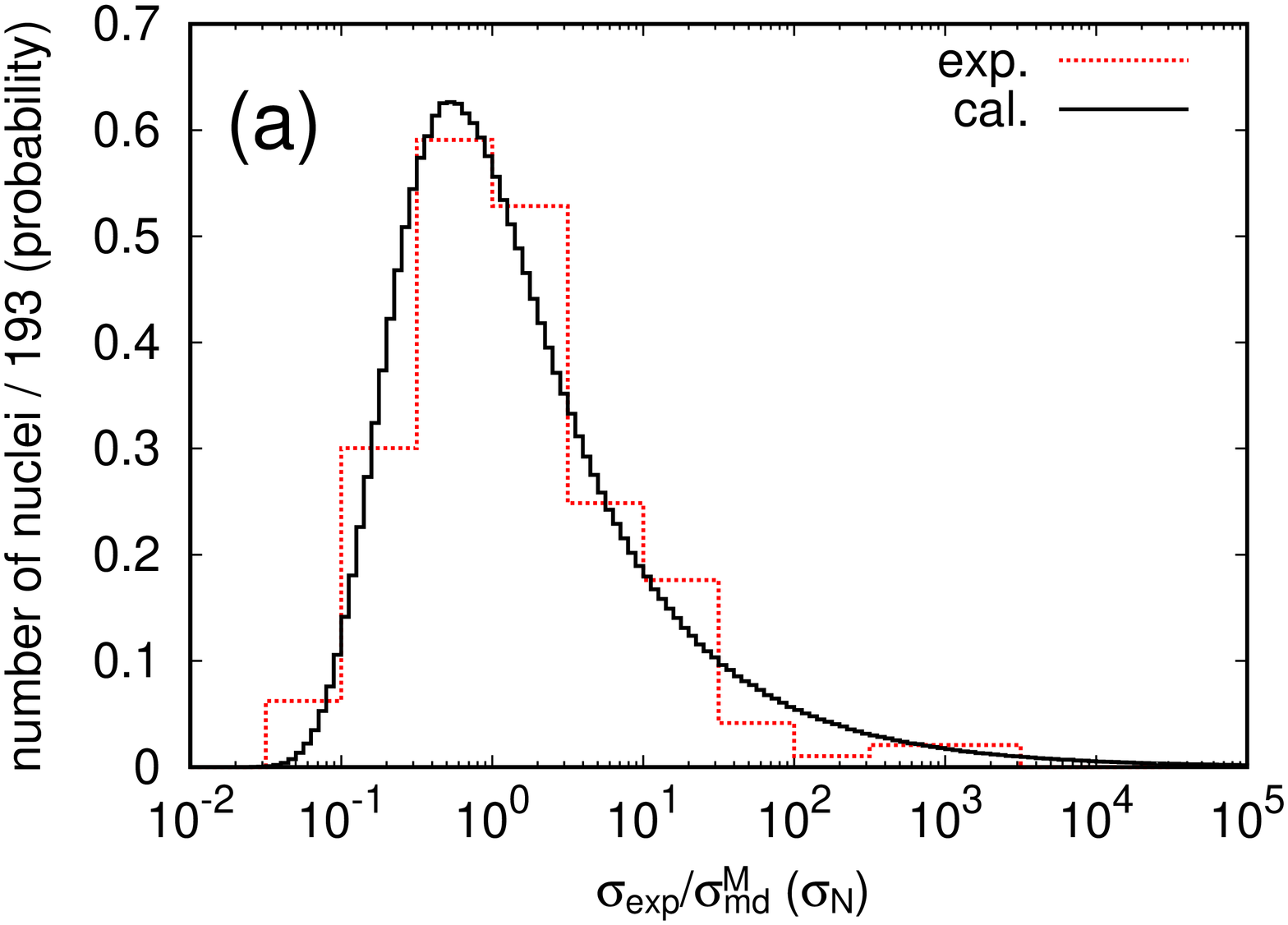} \\
	\includegraphics[scale=0.35,angle=270]{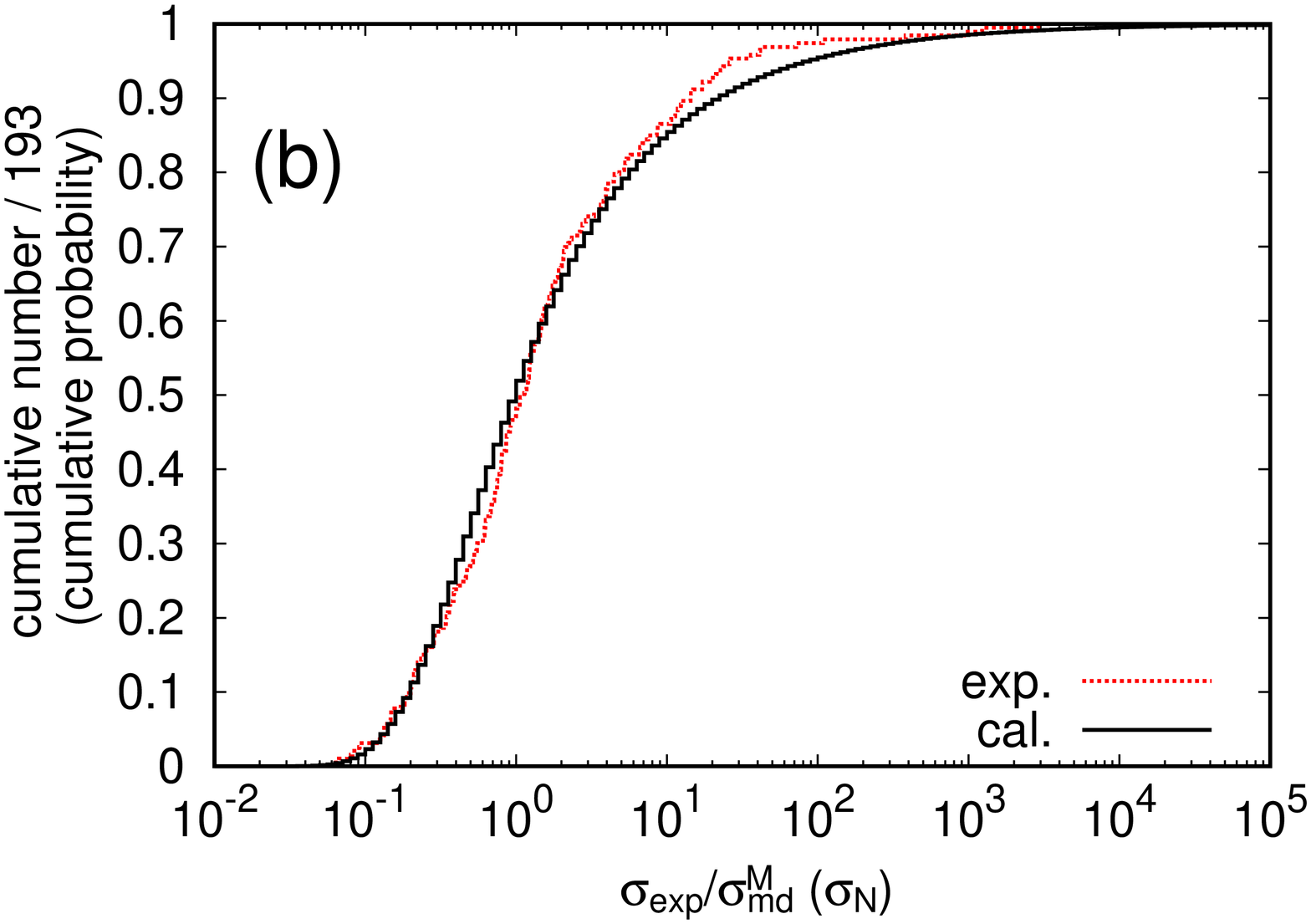}
	\caption{(a) Distribution ((b) cumulative distribution) of the nuclei as a function of the experimental $\sigma_{\rm th}$ normalized to $\sigma_{{\rm md}}^{\rm M}$ compared with $P_{\sigma}^{\rm M}\log (\sigma_{\rm N})$. The vertical axis is the number (cumulative number) of nuclei divided by the total number of nuclei, which are comparable to the probability (cumulative probability).} \label{fig6}
\end{figure}

In Fig. \ref{fig6} (a), $P_{\sigma}^{\rm M}(\log \sigma_{\rm N})$ with $D_0/\langle \Gamma_{\gamma 0} \rangle = 10^3$
is compared with the distribution of 193 nuclei as 
a function of the experimentally measured $\sigma_{\rm th}$ normalized to the calculated $\sigma_{\rm md}^{\rm M}$.  
It is noted that $P_{\sigma}^{\rm M}(\log \sigma_{\rm N})$ for some of 193 
nuclei in $D_0/\langle \Gamma_{\gamma 0} \rangle < 100$ eV 
region are varied from a typical distribution, as discussed above. 
While the tails of the probability in the large $\sigma_{\rm th}$ region slightly
shorten in these nuclei, such small variations are not important for the comparison with 
the distribution of the experimental data. 
Therefore, we also use the values of these nuclei for better statics.
We can see a fair agreement between $P_{\sigma}^{\rm M}(\log \sigma_{\rm N})$ 
and the distribution of the experimental data.
The agreement is even found in the extremely large $\sigma_{\rm th}$ region around 
$\sigma/\sigma_{{\rm md}}^{\rm M} \sim 10^3$ with four samples: $^{35}$Cl, $^{113}$Cd, $^{157}$Gd and $^{164}$Dy.

We also plot the cumulative probability in Fig. \ref{fig6} (b).
We can see a good agreement between the calculated cumulative probability and 
the cumulative number of the experimental data. 
This results support the validity of the major part of $P_{\sigma}^{\rm M}(\log \sigma_{\rm N})$.  

\section{Summary} \label{sum}
In this study, we investigated the probability distribution 
of the thermal neutron capture cross section $\sigma_{\rm th}$ 
calculated from the statistical properties of the resonance parameters.
In practice, the probability distribution was 
deduced numerically using the resonance parameters randomly sampled from Wigner and 
Porter-Thomas distributions. In the case of the single-resonance approximation, 
the analytical expression of the probability distribution was derived. 

We revealed that to what extent 
$\sigma_{\rm th}$ can be varied due to the statistical fluctuation of the resonance parameters from 
the probability distributions. 
It is important that the multi-resonance contribution significantly narrows the distribution 
and suppresses the emergence of a extremely small $\sigma_{\rm th}$.
Another important finding is that shapes of the probability distributions are very similar for nuclei that 
have $D_0$ sufficiently larger than $\langle \Gamma_{\gamma_0} \rangle$. 
The typical probability distribution for most of stable nuclei 
was compared with the distribution of the experimentally 
observed $\sigma_{\rm th}$. The validity of the present method was confirmed from a good agreement between them.

This study presents a fundamental knowledge to utilize the stochastic method to estimate 
$\sigma_{\rm th}$ for nuclei that have no available experimental data. The probability distribution can be 
used to evaluate the uncertainty in the calculated $\sigma_{\rm th}$. 
The methodology used in this study is not limited to the calculation of $\sigma_{\rm th}$, 
however will be useful to evaluate uncertainty arising from a statistical treatment of the resolved resonances in practical applications.

\begin{acknowledgments}
This work was funded by ImPACT Program of Council for Science, 
Technology and Innovation (Cabinet Office, Government of Japan).
\end{acknowledgments}

\appendix*

\section{}

Equation \eqref{eqf} is easily derived from the conversion of the random variables, 
which is well known for the normal and $\chi^2$ distributions.
We suppose that $E_1$ has Gaussian distribution with zero mean,
\begin{eqnarray}
P_{\rm se}(E_1)=\frac{1}{\sqrt{2\pi d}}\exp\left( -\frac{E_1^2}{2d} \right).
\end{eqnarray}
If we take the standard deviation $d=D_0^2/2\pi$, $P_{\rm se}(E_1)$ agrees with the
probability distribution of the first resonance numerically calculated using Wigner distribution.
By converting the random variable $E_1$ to $E_1^2$, it becomes $\chi^2$ distribution of one degree of freedom, 
\begin{eqnarray}
P_{\rm se}(E_1^2)=\frac{1}{\sqrt{2\pi d E_1^2}}\exp\left( -\frac{E_1^2}{2d} \right).
\end{eqnarray}
Then the probability distribution of $\Gamma_{n1}^{l=0}/E_1^2$ is calculated as 
\begin{eqnarray}
P_{\rm f}(Y) &=& \int_{0}^{\infty} P_w(YZ) P_{se}(Z) Z dZ , \\ \nonumber
                           &=& \frac{1}{\pi Y} \left( \sqrt{\frac{d}{\langle \Gamma_{n0}\rangle}Y}+\sqrt{\frac{\langle \Gamma_{n0}\rangle}{d}\frac{1}{Y}} \right)^{-1} , \\
Y &=& \Gamma_{n1}^{l=0}/E_1^2, \hspace{5mm} Z=E_1^2. \nonumber
\end{eqnarray}
The further conversion of the random variable $Y$ to $\sigma_{\rm th}=\frac{2}{k^2}\sqrt{E_0}\langle \Gamma_{\gamma 0}\rangle Y$
yields $P_{\sigma}^{\rm S}$ of Eq. \eqref{eqf}.
Equation \eqref{eqPf} is derived by replacing the exponential function emerged from the conversion of $\sigma_{\rm th}$ to $\log \sigma_{\rm th}$ 
\begin{eqnarray}
P_{\sigma}^{\rm S}(\log \sigma_{\rm th}) &=& \frac{\sqrt{\sigma_{{\rm md}}^{\rm S}}}{\pi}\left( {\rm e}^{\frac{\log \sigma_{\rm th}}{2}} 
+\sigma_{{\rm md}}^{\rm S} {\rm e}^{-\frac{\log \sigma_{\rm th}}{2}} \right)^{-1}, \nonumber
\end{eqnarray}  
with the hyperbolic functions. 

It is easily confirmed that $\sigma_{\rm md}^{\rm S}$ in Eq. \eqref{eqf} corresponds to the median of $P_{\sigma}^{\rm S}$ from the following calculations.
The cumulative distribution function of $\sigma_{\rm th}$ calculated form Eq. \eqref{eqf} is, 
\begin{eqnarray}
y = \int_0^X P_{\sigma}^{\rm S}(\sigma_{\rm th}) d\sigma_{\rm th} = \frac{2}{\pi}{\rm tan^{-1}}\sqrt{\frac{X}{\sigma_{{\rm md}}^{\rm S}}},
\end{eqnarray}
and its inverse function is, 
\begin{eqnarray}\label{eqcm}
X = \sigma_{{\rm md}}^{\rm S} {\rm tan^2} \left( \frac{\pi}{2}y \right).
\end{eqnarray}
By putting the cumulative probability $y=0.5$ in Eq.\eqref{eqcm}, we find $X=\sigma_{\rm md}^{\rm S}$.



\end{document}